\documentclass{article}
\usepackage{ITW06_PaperLaTeXStyle} 

\usepackage{amsmath}
\usepackage{amssymb}
\usepackage{color}
\usepackage{epsfig}
\usepackage{eepic}
\usepackage{graphicx}
\usepackage{psfrag}
\usepackage{cite}
\usepackage{epsf}
\usepackage{ifthen}
\usepackage{times}
\usepackage{amscd}
\usepackage{amsfonts}
\usepackage{bm}
\usepackage{multicol}

\begin{document}
\itwAfour             % Uncomment for A4 paper
\topmargin = 0mm

\itwtitle{Non-binary Hybrid LDPC Codes: structure, decoding and optimization}

\itwauthorswithsameaddress{Lucile Sassatelli and David Declercq}
{ETIS - ENSEA/UCP/CNRS UMR-8051\\
95014 Cergy-Pontoise, France \\
{\tt \{sassatelli, declercq\}@ensea.fr}}

\itwmaketitle

\footnotetext[1]{This work was supported by the French Armament Procurement Agency (DGA).}

\begin{itwabstract}
In this paper, we propose to study and optimize a very general class of LDPC codes whose variable nodes belong 
to finite sets with different orders. We named this class of codes {\em hybrid LDPC codes}. 
Although efficient optimization techniques exist for binary LDPC codes and more recently for non-binary 
LDPC codes, they both exhibit drawbacks due to different reasons. Our goal is to capitalize on the advantages 
of both families by building codes with binary (or small finite set order) and non-binary parts 
in their factor graph representation. The class of hybrid LDPC codes is obviously larger than existing 
types of codes, which gives more degrees of freedom to find good codes where the existing codes show 
their limits. We give two examples where hybrid LDPC codes show their interest.
\end{itwabstract}

\begin{itwpaper}

%%%%%%%%%%%%%%%%%%%%%%%%%%%%%%%%%%%%%%%%%%%%%%%%%%%%%%%%%%%%%%%%%%%%%%%%%%%%%%%%%%%%%%%%%%%%%%%%%%%%%%%%%%%%%%%%%%%%%
\itwsection{Introduction}
Binary LDPC codes are now well recognized as capacity approaching codes for various types of channels when the size 
of the codeword tends to infinity, and various methods have been proposed to optimize their irregularity profile with the 
help of Density Evolution under Gaussian Approximation (DE-GA) \cite{Chung}. Other techniques based on EXIT charts \cite{tenBrink} 
are also related to DE-GA and lead to the same analysis and optimization algorithms. However, there are several issues for 
which the binary LDPC codes show their limits: we can cite for example coded modulations and/or coding for small or moderate 
block lenghts. For these contexts, it has been shown recently that non-binary LDPC codes can be a good alternative. They exhibit 
better performance than their binary counterparts for coded modulations \cite{Bennatan} and for code length typically in the 
range $N\in[500,2000]$ information bits \cite{hu:icc04,PoulliatTC2006}. The main interest of non-binary LDPC codes actually lies in 
the decoder: good non-binary LDPC codes have much sparser factor graphs (or Tanner graphs) than binary LDPC codes \cite{davey:cl98}, 
and the Belief Propagation (BP) decoder is closer to optimum decoding since the small cycles can be avoided with a proper graph 
construction, as proposed in \cite{hu:icc04}. In this paper, we propose to study a class of hybrid LDPC codes which aims at combining the advantages of binary and non-binary 
LDPC codes in the same coding scheme. The class of hybrid LDPC codes is a generalization of existing classes of LDPC codes. For 
hybrid LDPC codes, we allow the connectivity profile of the factor graph to be irregular, but also the order of the symbols in a 
codeword can be irregular, that is to say, the symbols can belong to finite sets with different orders. We depict in section 
II the structure of hybrid LDPC codes and briefly describe the decoding algorithm. In section III, we recall the existing work 
on optimization of non-binary LDPC codes with DE-GA, and introduce a specific modelization of the messages in the factor graph 
which allow an efficient optimization of non-binary LDPC codes on the binary input Gaussian channel (BI-AWGN). In section IV, the 
DE-GA equations for hybrid LDPC codes are derived and the optimization procedure is presented. The analysis of hybrid LDPC codes 
is based on a detailed representation of the factor graph of the code \cite{Kasai}, together with the introduction of extra 
parameters to describe the proportion of irregular set orders in the codeword. The parameterization of hybrid LDPC codes is therefore 
very rich. We have then decided to optimize sub-classes of hybrid LDPC codes, and we give two different examples that show 
their interest when compared to the best known existing LDPC codes. The examples and the simulation results are shown in section V.

%%%%%%%%%%%%%%%%%%%%%%%%%%%%%%%%%%%%%%%%%%%%%%%%%%%%%%%%%%%%%%%%%%%%%%%%%%%%%%%%%%%%%%%%%%%%%%%%%%%%%%%%%%%%%%%%%%%%%
\itwsection{The class of Hybrid LDPC Codes}
We define a non-binary hybrid LDPC code as an LDPC code whose variable nodes belong to finite sets of different orders. 
To be specific, this class of codes is not defined in a finite field, but in finite groups.
We will only consider groups whose cardinality $q_k$ is a power of $2$, that says groups of the type $G(q_k)=\left(\frac{\mathbb{Z}}{2\mathbb{Z}}\right)^{p_k}$ with
$p_k=\log_2(q_k)$. Thus each element of $G(q_k)$ has a binary map of $p_k$ bits. Let us call the minimum order of codeword symbols $q_{min}$, and the maximum order of codeword symbols $q_{max}$. The class of hybrid LDPC codes is defined on the product group $\left(\frac{\mathbb{Z}}{2\mathbb{Z}}\right)^{p_{min}}
\times\hdots\times\left(\frac{\mathbb{Z}}{2\mathbb{Z}}\right)^{p_{max}}$.
Note that this type of LDPC codes built on product groups has already been proposed in the literature \cite{Sridhara2002}, but no optimization of the code structure has 
been proposed and its application was restricted to the mapping of the codeword symbols to different modulation orders. Parity check codes defined on $\left(G(q_{min})\times \ldots \times G(q_{max})\right)$ are particular since they are linear in $\frac{\mathbb{Z}}{2\mathbb{Z}}$, 
but could be non-linear in the product group. Although it is a loss of generality, we have decided to restrict ourselves to hybrid LDPC codes that 
are linear in their product group, in order to bypass the encoding problem. We will therefore only consider upper-triangular parity check matrices and a 
specific ordering of the symbol orders in the codeword, which ensures the linearity of the hybrid codes. The structure of the codeword and the 
associated parity check matrix is depicted in Figure \ref{H}.
\begin{figure}[!h]
\centering
\input{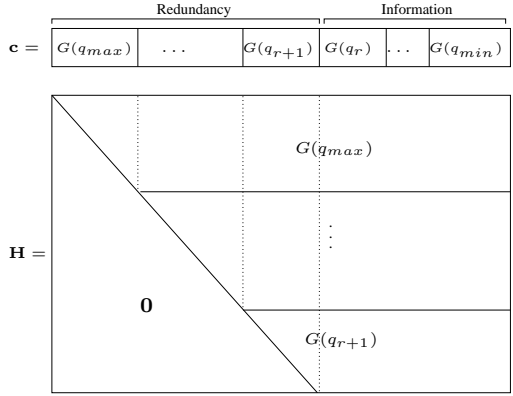}
\caption{Hybrid codeword and parity-check matrix.}
\label{H}
\vspace*{-0.5cm}
%\vspace*{0.5cm}
\end{figure}
We hierarchically sort the different group orders in the rows of the parity-check matrix, and also in the codeword, such that 
$q_{min}<\ldots<q_{k}<\ldots<q_{max}$. To encode a redundancy symbol, we consider each symbol that participates in the parity check 
as an element of the highest group, which is only possible if the groups are sorted as in Figure \ref{H}. This clearly shows that encoding 
is feasible in linear time by backward computation of the check symbols.\\
In order to explain the decoding algorithm for hybrid LDPC codes, it is usefull to interpret a parity check the hybrid code as a special case of a parity check 
built on the highest order group of the symbols of the row, denoted $G(q_l)$ and have a look at the binary image of the equivalent code \cite{PoulliatTC2006}. For codes defined 
over Galois fields, the nonzero values of $H$ correspond to the companion matrices of the finite field elements and are typically rotation matrices (because of the 
cyclic property of the Galois fields). In the case of hybrid LDPC codes, the nonzero values have no linear representation and are indeed nonlinear maps that have rectangular matrix equivalents. To be more specific, the function that connects a row in $G(q_l)$ and a column in $G(q_k)$ is a nonlinear function that maps 
the $q_k$ symbols of $G(q_k)$ into a subset of $q_k$ symbols that belongs to $G(q_l)$. This function has an equivalent binary representation by a matrix of 
dimension $(p_l \times p_k)$. Note that with the above mentioned constraints, we have necessarily $p_k<p_l$. It is not very difficult to generalize the 
Belief propagation decoder to hybrid codes, and it has been shown that even for those very particular structures, it is possible to derive a fast version of the decoder using FFTs \cite{Goupil}. For lack of space reason, we do not present in this paper the BP decoder for hybrid codes, and we refer to the general algorithm described in \cite{Goupil} 
for which the decoder for hybrid LDPC codes is a special case. In the rest of the paper, we will call the message passing step through $h_{ij}$ {\em extension} when it is from $G(q_k)$ to $G(q_l)$ and 
{\em truncation} when it is from $G(q_l)$ to $G(q_k)$ since the sizes for the messages in the factor graph differ. The BP decoder steps can be 
followed in the factor graph representation of a single parity check as depicted in Figure \ref{dec}.
\begin{figure}[!h]
\centering
\input{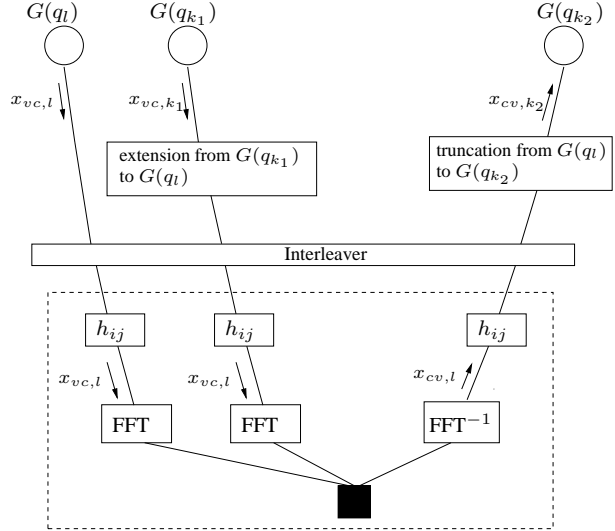}
\caption{A $G(q_l)$ check node of a hybrid decoder.}
\label{dec}
\end{figure}
%%%%%%%%%%juste pour mettre equations sur la bonne page, ne va pas ds cette section%%%%%%%%%%%%%%%%%%%%%%%%%%%%%%%%%
% \begin{figure*}[!htb]
% \begin{equation}\label{mipure}
% x_{vc}^{(t+1)}=\sum_i \lambda_i J_v\left(\mathbf{m}_{sc}+(i-1){J_c}^{-1}\left(1-\sum_j \rho_j J_c\left((j-1){J_c}^{-1}(1-x_{vc}^{(t)})\right)\right)\mathbf{1}_{q-1}\right)
% \end{equation}
% \end{figure*}
%%%%%%%%%%%%%%%%%%%%%%%%%%%%%%%%%%%%%%%%%%%%%%%%%%%%%%%%%%%%%%%%%%%%%%%%%%%%%%%%%%%%%%%%%%%%%%%%%%%%%%%%%%%%%%%%%%%%%
Let us now introduce parameters that describes the irregularity of group orders in the codeword. 
Let $\tilde{\gamma}_k$ be the proportion of symbol nodes in the hybrid graph which belong to $G(q_k)$ 
and by definition, we take $q_{min}=2$. The code rate of an hybrid code with the specific structure 
presented in Figure \ref{H} can be expressed as:
\[ R=\frac{\sum_{k=min}^{r} \tilde{\gamma}_k log_2(q_k)}{\sum_{k=min}^{max} \tilde{\gamma}_k log_2(q_k)} \] 
Note that this expression is completely general since if we fix $q_r=q_{r+1}$, then both information 
and redundancy can share the same group order $q_r$. 
In order to optimize hybrid LDPC codes, following the strategies used to optimize binary or non-binary 
LDPC codes, we need to express the density evolution of the messages under Gaussian approximation 
along one decoding iteration, with respect to the parameters to be optimized. In our case, the parameters are 
the proportions of irregular connections in 
the graph {\em and} the proportions of irregular group orders $\tilde{\gamma}_k$. In the next section, 
we recall some required properties of DE-GA for non-binary LDPC codes, that we will use to make 
the generalization to hybrid codes.

\itwsection{Analysis of Non-binary LDPC Codes over GF$(q)$}
%%%%%%%%%%juste pour mettre equations sur la bonne page, ne va pas ds cette section%%%%%%%%%%%%%%%%%%%%%%%%%%%%%%%%%
\begin{figure*}[!htb]
{\small
\begin{equation}\label{mipure}
x_{vc}^{(t+1)}=\sum_i \lambda_i J_v\left(\mathbf{m}_{sc}+(i-1){J_c}^{-1}\left(1-\sum_j \rho_j J_c\left((j-1){J_c}^{-1}(1-x_{vc}^{(t)})\right)\right)\mathbf{1}_{q-1}\right)
\end{equation}
% \begin{equation}\label{mihybi}
% x_{cv,q_s}^{(t-1)}=1-\sum_j \rho_j J_c((j-1){J_c}^{-1}(1-x_{vc,q_s}^{(t-1)},q_s),q_s)
% \end{equation}
% \begin{equation}\label{mihyb}
% x_{vc,q_s}^{(t)}=\sum_i \sum_{k=min}^{s}\pi(i,k)\left(1-\frac{log(q_k)}{log(q_s)}\left(1-J_v\left(\mathbf{m}_{sc}^{q_k}+(i-1){J_c}^{-1}(x_{cv,q_s}^{(t-1)},q_s)\mathbf{1}_{q-1},q_k\right)\right)\right)
% \end{equation}}
\begin{equation}\label{mihybi}
{x_{cv,k}^{(j,l')}}^{(t)}=J_c\left({J_c}^{-1}\left(1-J_c\left((j-1){J_c}^{-1}(1-\sum_{i',k'}\pi(i',k'|j,l'){x_{vc,l'}^{(i',k')}}^{(t)},q_{l'}),q_{l'}\right),q_{l'}\right),q_k\right)
\end{equation}
\begin{equation}\label{mihyb}
{x_{vc,l}^{(i,k)}}^{(t)}=1-\frac{log(q_k)}{log(q_l)}\left(1-J_v\left(\mathbf{m}_{sc}^{q_k}+(i-1){J_c}^{-1}(\sum_{j,l'}\pi(j,l'|i,k){x_{cv,k}^{(j,l')}}^{(t-1)},q_k)\mathbf{1}_{q-1},q_k\right)\right)
\end{equation}}
\end{figure*}
%%%%%%%%%%%%%%%%%%%%%%%%%%%%%%%%%%%%%%%%%%%%%%%%%%%%%%%%%%%%%%%%%%%%%%%%%%%%%%%%%%%%%%%%%%%%%%%%%%%%%%%%%%%%%%%%%%%%%
Let us first present how one analyses non-binary LDPC codes on the BI-AWGN channel using a gaussian approximation. We must quote three works from which this approach is highly inspired. In \cite{Bennatan}, Bennatan et al. have proposed a density evolution for LDPC codes on $GF(q)$ on memoryless $q$-ary channels. They have derived a gaussian approximation of the densities of messages, which leads to a quite easy optimization of these codes, using EXIT charts \cite{tenBrink}. Although very general, their approach can be improved if 
the channel is BI-AWGN, by choosing a more accurate initialization of the densities of the LLR messages. 
In our work, we took a different initialization of the decoder, which describes more precisely the 
BI-AWGN output messages. The model for the LLRs is the same as the one proposed in \cite{Takawira}, where 
the authors analyse non-binary LDPC codes on the BI-AWGN channel. Following well known ideas, we will track 
the information content of the messages (here vector messages) under Gaussian approximation, that is 
the mutual information of a discrete input channel with additive Gaussian noise whose output is the message 
in the graph. In \cite{Li}, Li. et al. have also proposed a DE-GA approach for non-binary LDPC codes, 
but the quantity they used to follow the evolution of densities was the mean of the messages instead of the mutual information.
The first necessary property that must be fulfilled is a symmetry property for the vector messages. 
The symmetry of $q$-ary log density ratio (LDR) vector $\mathbf{W}$ is defined in \cite{Bennatan}. 
Let $v$ be the corresponding symbol sent and $W_a$ the $a$-th component of $\mathbf{W}$. An LDR-vector is symmetric if and only if $\mathbf{W}$ satisfies 
\begin{equation}\label{symmBen}
p(\mathbf{W}|v=a)=e^{-W_a}p(\mathbf{W}|v=0), \qquad \forall a \in GF(q)
\end{equation}
In \cite{Bennatan}, the symmetry was defined for codes defined on fields, but this definition clearly 
applies for finite Abelian groups. For the BI-AWGN channel, the bitwise log likelihood ratios (LLR) 
are symmetric in the sense defined in \cite{Chung}, which in turn, induces the symbolwise symmetry of the 
LLR-vector. Moreover, the symmetry property (\ref{symmBen}) is kept during the non-binary 
BP decoder operations \cite{Li}. In the next section, we discuss the compatibility of the symmetry 
property with the specific operations used in the hybrid decoder, that is the truncations and extensions. 
We now define some useful notations, in concordance with the previous quoted works, to express information 
transfer functions. $LLR_b$ denotes the  bitwise LLR of a received BPSK modulated bit and $m_{bc}$ is the mean
of $LLR_b$. $LLR_s$ denotes a symbolwise LLR vector of a $GF(q)$ symbol and $\mathbf{m}_{sc}$ its vector mean.
If $\mathbf{B}$ is the $(q-1)\times p$ (with $p=log_2(q)$) mapping matrix from vectors of $p$ bits to $GF(q)$ 
symbols and $\mathbf{1}_p$ is the all-one column vector of size $p$, then we have 
$\mathbf{m}_{sc}=m_{bc}\mathbf{B}\mathbf{1}_p$. 
If we call $\sigma^2$ the variance of the BI-AWGN channel, thanks to the symmetry of the channel, we know from \cite{Chung} that $m_{bc}=\frac{2}{\sigma^2}$ and $LLR_b \sim {\mathcal N}(m_{bc},2m_{bc})$. As said previously,  the symbolwise LLRs are then symmetric. According to \cite{Li}, if the messages are symmetric and gaussian  distributed as ${\mathcal N}(\mathbf{m},\mathbf{\Sigma})$, the covariance matrix $\mathbf{\Sigma}$ can be uniquely determined by the mean vector $\mathbf{m}$ such that
\begin{equation}\label{symm}
\mathbf{\Sigma}_{i,j}=\mathbf{m}_i+\mathbf{m}_j-\mathbf{m}_{i\oplus j},\qquad i,j \in GF(q)
\end{equation}
Again, this property is defined in a Galois field, but remains the same in a group of order q since 
it only requires the use of the proper addition $\oplus$ in the Abelian group. 
The symmetry allows to make the all-zero codeword assumption. If we make the approximation that all the vector messages on the graph are gaussian, then we can see on Equation (\ref{symm}) that we need to track only the $(q-1)$ components of the mean vector to get full-description of the densities. If the nonzeros values in the 
parity matrix $H$ are choosen uniformly, it follows that the components of the mean vector of any check node outcoming message are equal to the same scalar $m_{cv}$. 
The mean vector of LDR-vectors going out from data nodes is entirely determined by the variance of 
the BI-AWGN channel, the mapping $\mathbf{B}$, and the mean of check node outgoing LDR-vectors. 
Combining all these results, one can show \cite{Li,Bennatan} that only two scalar parameters entirely 
define the gaussian approximation of densities of messages on the graph: $\sigma^2$ and $m_{cv}$. 
Since the channel is known at each step of the optimization process, only one scalar parameter remains to 
track: $m_{cv}$. Using the one-to-one relation between the scalar mean of a vector and its mutual information
given in equation \ref{defmi}, we can express the EXIT transfer function of one iteration of the non-binary 
BP decoder.
\begin{equation}\label{defmi}
I_{\mathbf{v}}=1-\mathbb{E}_{\mathbf{v}}\left(log_q(1+\sum_{i=1}^{q-1} e^{-v_i})\right)
\end{equation}
Let us denote the two useful functions $J_v$ and $J_c$ (for variable node decoder and check node decoder, respectively), determined as in \cite{Takawira}:
\begin{eqnarray}
J_v(\mathbf{m})&=&1-\mathbb{E}_{\mathbf{v}}\left(log_q(1+\sum_{i=1}^{q-1} e^{-v_i})\right),\\ &&\mbox{ with }\mathbf{v}\sim {\mathcal N}( \mathbf{m},\mathbf{\Sigma})\nonumber\\
J_c(m)&=&1-\mathbb{E}_{\mathbf{v}}\left(log_q(1+\sum_{i=1}^{q-1} e^{-v_i})\right),\\ &&	\mbox{ with }\mathbf{v}\sim {\mathcal N}(m\mathbf{1}_{q-1},\mathbf{\Sigma})\nonumber
\end{eqnarray}
where $\mathbf{\Sigma}$ is computed from $\mathbf{m}$ by the symmetry relation of Equation (\ref{symm}). 
Note that $J_c$ is a particular case of $J_v$ where all components of the vector $\mathbf{m}$ are equal to $m$. Finally, we get Equation (\ref{mipure}) that expresses the extrinsic transfer 
function of the non-binary BP decoder used on a BI-AWGN channel from iteration number $t$ to iteration number $t+1$. $(\lambda,\rho)$ are the usual parameters 
that define the connectivity profile of a family of $GF(q)$ LDPC codes, and $x_{vc}^{(t)}$ is the mutual information of any check node incoming vector messages at the $t$-th iteration. For more details 
about the derivation of these equations and the associated proofs, please refer to the cited papers.

\itwsection{Analysis of Hybrid LDPC Codes}
In this section, we now explain how we can generalize the equations of DE-GA of non-binary LDPC codes 
to hybrid LDPC codes and how to introduce the extra parameters that describe the irregularity in the 
group orders. To properly define a family of hybrid codes, it is usefull to adopt a detailed 
representation of the factor graph, directly inspired from the one introduced by Kasai et al. 
in \cite{Kasai}. We define a hybrid LDPC code family by $\pi(i,j,k,l)$. It is the joint probability that an edge of the hybrid Tanner graph is linked to a data 
node of connectivity degree $i$ in $G(q_k)$ and to a check node of connectivity degree $j$ in $G(q_l)$.
We also define the following marginal and conditionnal probabilities
{\small
\[ \gamma_k=\sum_{l=min}^{max}\sum_{i,j} \pi(i,j,k,l)\qquad,\qquad \lambda_i=\sum_{k,l=min}^{max} \sum_{j}\pi(i,j,k,l) \]}
{\tiny
\[ \lambda(i,k)=\frac{\sum_{l=min}^{max}\sum_{j} \pi(i,j,k,l)}{\gamma_k}\qquad,\qquad \gamma(i,k)=\frac{\sum_{l=min}^{max}\sum_{j} \pi(i,j,k,l)}{\lambda_i} \]
}
$\lambda(i,k)$ is the proportion of edges linked to a symbol node of degree $i$, given that this symbol 
node is in $G(q_k)$, and $\gamma(i,k)$ is the proportion of edges linked to a symbol node in $G(q_k)$, 
given that this symbol node is of degree $i$. 
The analysis of hybrid non-binary LDPC codes is completely based on the previous approach that assumes the densities of vector messages to be gaussian distributed, when transmitting on BI-AWGN channel. We add two steps to the non-binary analysis described in the last section, that correspond to truncation and extension 
of messages when passing from a data node to a check node in a higher order group, and vice versa.
Thanks to Equation (\ref{defmi}), we easily obtain the expression between the mutual information 
$x_{q_l}$, of an extended LDR message in $G(q_l)$, built from a message in $G(q_k)$ whose mutual information is $x_{q_k}$:
\[(1-x_{q_k})log_2(q_k)=(1-x_{q_l})log_2(q_l)\]
To get the relation giving the mutual information of a message in $G(q_k)$ built by the truncation of an 
LDR message in $G(q_l)$, we need to redefine the functions $J_v(\mathbf{m})$ and $J_c(m)$. 
$J_v(\mathbf{m},q)$ and $J_c(m,q)$ are defined in the same way as before, with $q$ that represents 
the order of the group of the vector messages whose mean is $\mathbf{m}$ or $m\mathbf{1}_{q-1}$. 
With these new definitions of functions $J_v$ and $J_c$, if $x_{q_k}$ is the mutual information 
of the truncated vector, we have:
\[x_{q_k}=J_c({J_c}^{-1}(x_{q_l},q_l),q_k)\]
which corresponds to the conservation of the mean of each component after truncation. We also re-define $\mathbf{m}_{sc}$ by $\mathbf{m}_{sc}^q$ where $q$ is the order of the group of the symbol node 
whose LLR vector of size $q-1$ has mean $\mathbf{m}_{sc}^q$. We have also shown that the symmetry property 
of the messages holds for the specific transformations of truncation and extension. We do not present the 
proofs here and they will be reported in future publication. Following the different steps of one decoding 
iteration, we can derive the EXIT function of one iteration of the hybrid decoder. This EXIT function 
is expressed in equations (\ref{mihybi}) and (\ref{mihyb}). 
This function expresses ${x_{vc,l}^{(i,k)}}^{(t)}$ (resp. ${x_{cv,k}^{(j,l)}}^{(t)}$) which is the mutual information at the $t$-th iteration of a vector message going out of a data (resp. check) node of degree $i$ (resp. $j$) in $G(q_k)$ (resp. $G(q_l)$) extended (resp. truncated) to become input of a check (resp. data) node in $G(q_l)$ (resp. $G(q_k)$).

\itwsection{Optimization and Results}
As for the optimization of usual LDPC codes, we want to find the parameters of a hybrid family for a given 
code rate that minimize the convergence threshold. In all the simulations presented in this paper, we have considered all the check nodes of same degree and in the same group. And hence the number of parameters to be optimized is reduced from four to two (distribution $\pi(i,k)$ of degrees and groups of variable nodes). The ideal optimization procedure would be to jointly optimize $\gamma$ and $\lambda$, i.e., the 2-variable function $\pi(i,k)$.
In order to simplify the optimization, we chose to fix one of these parameters, and to optimize 
the other one. That says we tested two directions of optimizing hybrid LDPC codes: 
either we look for the optimal proportions $\gamma_k$ of different finite sets given a fixed connectivity 
of the graph $\lambda_i$, or we look for the optimal proportions $\lambda_i$ given a fixed repartition $\gamma_k$ of the group orders in the codeword. 
For both approaches, we choose to map all the redundancy bits into symbols in the highest order 
group $G(q_{max})$, and to prohibit information symbol nodes that are in $G(2)$ to be of degree 2 in order 
to mitigate the influence of catastrophic cycles. 
First, we consider the optimization of $\lambda$, when $\gamma(i,k)$ is fixed. 
From the above remarks, it follows that we fix as a priori constraints 
$\gamma(2,2)=0,\gamma(i\neq 2,2)=1$.
The other parameters $\gamma(2,q)$ for $q\neq 2$ are determined by the proportions 
of information symbols in the different groups. For this simplified model, the code rate is 
defined by:
\[ R=1-\frac{\sum_j \frac{\rho_j}{j} log_2(q_{max})}{\sum_i  
	\frac{\lambda_i}{i}\sum_{k=min}^{max}\gamma(i,k)log_2(q_k)}\]
According to this expression, the code rate maximization is equivalent to the maximization of the denominator of the second term. 
Moreover, since $\pi(i,k)=\lambda_i \gamma(i,k)$, equation (\ref{mihyb}) corresponds to the 
convergence criterion equivalent to a strictly increasing information content 
$x_{vc,q_{max}}^{(t+1)}>x_{vc,q_{max}}^{(t)}$. Thus the cost function and all constraints 
are linear with respect to $\lambda$ and the optimization problem can be efficiently 
solved using linear programming.\\
The hybrid code solution of the optimization problem is relatively 
dense since it has an average row weight of 14.3 ones, but it comes from the fact that a rate 1/2 hybrid 
code is obtained with a graph with higher rate. Indeed, the hybrid LDPC codes are adapted for rather 
low rates. In Figure \ref{fig:sim1}, we give the simulation results for a code with target rate $R=1/2$. 
The hybrid code is compared to existing good codes. The irregular binary code has been chosen from 
the distributions in \cite{Chung} and the distribution of the irregularity for the $GF(8)$ code has 
been optimized with the equations of section III. All graphs have been designed with the PEG algorithm 
that has been widely accepted as a good finite length code construction.
\begin{figure}
\centering
\includegraphics[width=0.43\textwidth]{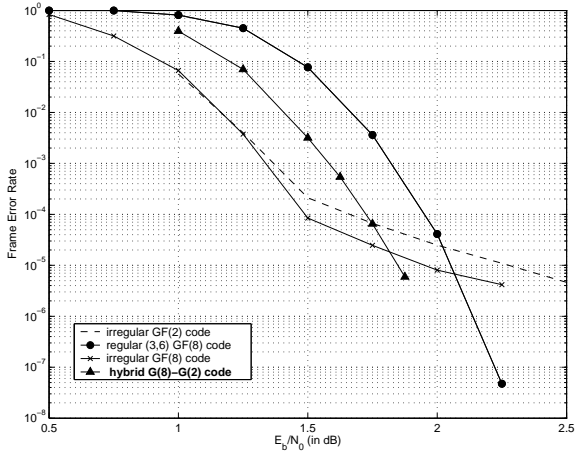}
\caption{Frame Error Rate comparison of hybrid LDPC codes with other good codes, $R=1/2$, $N_{bit}=3008$, 
the maximum number of iterations has been fixed to 500.}
\label{fig:sim1}
\end{figure}
First, we can see that the error floor is lowered by going from $GF(2)$ to $GF(8)$, and that the 
regular $(3,6)$ code in $GF(8)$ has a worse convergence than the irregular codes, but a much lower 
error floor. Those results are in accordance with the usual observations on binary LDPC codes. Our 
hybrid LDPC code with 2 group orders $G(8)-G(2)$ is as expected a good compromise of the joint 
problem convergence/error floor. The convergence region has been slightly degraded compared to 
irregular LDPC codes, but with the effect of no observed error floor up to a FER=$5.10^{-6}$. We 
expect even better results by allowing more degrees of freedom in the optimization procedure.\\
In the second example, we optimized $\gamma_k$, with $\lambda(i,k)$ fixed. In this case, 
we look for the best proportion of group orders for a regular hybrid graph defined by the 
connectivity of data nodes and check nodes $(d_v=2,d_c=3)$. According to the code rate expression
\[ R=1-\frac{\frac{1}{d_c} log_2(q_{max})}{\frac{1}{d_v}\sum_{k=min}^{max}\tilde{\gamma}_k log_2(q_k)} \]
the cost function is still the denominator of the second term. We aimed with this example at 
designing good codes for a rather low rate of R=1/6. We obtained the optimized hybrid code with 
three different group orders $G(256)-G(16)-G(8)$, and we have compared our hybrid code with various 
good codes presented in the literature. In Figure \ref{fig:sim2}, we can see that the irregular 
binary LDPC code is not a good solution for such low rate and moderate block length, as it is the 
worst code simulated. The regular code over $GF(256)$ designed with the methods presented in 
\cite{PoulliatTC2006} is better with $0.5dB$ gain, but is outperformed by a very specific 
construction of binary quasi-cyclic LDPC codes especially designed for low rates found in \cite{Ryan}. 
Our hybrid code shows the best performance and is to our knwoledge the best performance observed 
at this rate and codelengths. This confirms the fact that hybrid LDPC codes appear to be a good 
solution for low rate applications. Note that the error floor of our hybrid code is likely to 
be lowered with similar techniques as presented in \cite{PoulliatTC2006}. We plan to address this issue 
in a future work.
\begin{figure}
\centering
\includegraphics[width=0.43\textwidth]{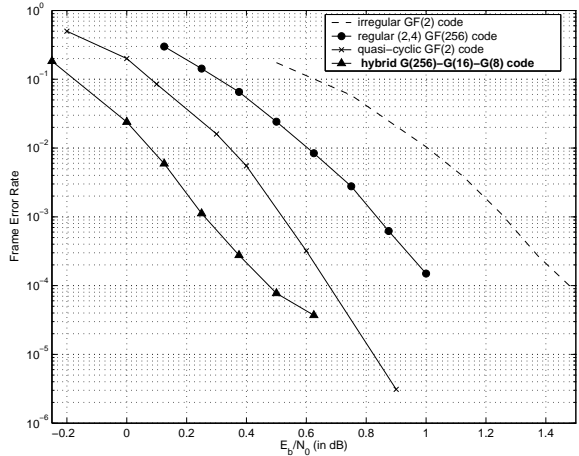}
\caption{Frame Error Rate comparison of hybrid LDPC codes with other good codes, $R=1/6$, $N_{bit}=6144$, 
the maximum number of iterations has been fixed to 500.}
\label{fig:sim2}
\end{figure}

\itwsection{Conclusion}
This paper aims at combining advantages of having variable nodes in different order finite sets, in a bipartite graph, to build non-binary hybrid LDPC codes. First, we have presented the structure and the decoding of the class of hybrid codes. We have then explained how to optimize irregular non-binary LDPC codes over GF$(q)$ for the BI-AWGN channel, and we have described how to generalize this technique for the optimization of hybrid codes. Finally, the most interesting results are obtained for quite low target code rates ($R=1/6$): our hybrid code outperforms the best known codes for this code rate. Future work will address the problem of the finite-length optimization for this class of codes.

% \itwacknowledgments
% 
% The authors would like to thank various sponsors for supporting
% their research.

\end{itwpaper}

\begin{itwreferences}

% \bibitem{Shannon1948}
% C. E. Shannon, ``A mathematical theory of communication,''
% \emph{Bell Syst.\ Tech.\ J.}, vol.\ 27, pt.~I, pp.~379--423, 1948;
%      pt.~II, pp.~623--656, 1948.

% \bibitem{Gallager62}
% R.G. Gallager, ''Low-Density Parity-Check Codes,''
% \emph{IRE Trans. on Inform. Theory}, pp. ~21--28, 1962.

\bibitem{Kasai}
K. Kasai, T. Shibuya and K.Sakaniwa, ''Detailedly Represented Irregular LDPC Codes,''
\emph{IEICE Trans. Fundamentals}, vol. E86-A(10) pp. 2435-2443, Oct. 2003.

%\bibitem{Urbanke2001}
%T. Richardson, A. Shokrollahi and R. Urbanke, ''Design of Capacity-Approaching Irreglar LDPC Codes,''
%\emph{IEEE Trans. on Inform. Theory}, vol.\ 47, no. 2, pp. 619-637, Feb. 2001.

\bibitem{Chung}
S.Y. Chung, T. Richardson and R. Urbanke, ''Analysis of Sum-Product Decoding LDPC Codes using a Gaussian Approximation,''
\emph{IEEE Trans. on Inform. Theory}, vol.\ 47, no. 2, pp. 657-670, Feb. 2001.

%\bibitem{Declercq2005}
%D. Declercq and M. Fossorier, ''Decoding Algorithms for Nonbinary LDPC Codes over GF(q),''
%\emph{sub. to the IEEE Trans. on Comm., April 2005}

\bibitem{Bennatan}
A. Bennatan and David Burshtein, ''Design and Analysis of Nonbinary LDPC Codes for Arbitrary Discrete-Memoryless Channels,''
\emph{IEEE Trans. on Inform. Theory}, vol.\ 52, no. 2, pp. 549-583, Feb. 2006.

\bibitem{Takawira}
G. Byers and F. Takawira, ''EXIT Charts for Non-binary LDPC Codes,''
\emph{Proc. IEEE ICC 2005}, vol.\ 1, pp. 652-657, May 2005.

\bibitem{Li}
G. Li, I. Fair and W. Krzymien, ''Analysis of Nonbinary LDPC Codes Using Gaussian Approximation,''
\emph{Proc. IEEE ISIT 2003}, p. 234, July 2003.

\bibitem{tenBrink}
S. ten Brink, ''Convergence Behavior of Iteratively Decoded Parallel Concatenated Codes,''
\emph{IEEE Trans. on Comm.}, vol.\ 49, no. 10, pp. 1727-1737, Oct. 2001.

\bibitem{hu:icc04}
X.-Y. Hu and E. Eleftheriou, 
``Binary Representation of Cycle Tanner-Graph {GF$(2^q)$} Codes,''
{\em The Proc. IEEE Intern. Conf. on Comm.}, Paris, France, pp. 528-532,
June 2004.

\bibitem{PoulliatTC2006}
C. Poulliat, M. Fossorier and D. Declercq, ``Using Binary Image of Nonbinary LDPC Codes to Improve Overall Performance,''
{\em in IEEE Intern. Symp. on Turbo Codes}, Munich, April 2006.

\bibitem{Goupil}
A. Goupil, M. Colas, G. Gelle and D. Declercq, 
``FFT-based BP Decoding of General LDPC Codes over Abelian Groups,''
\emph{to appear in IEEE Trans. on Comm.}, 2006.

\bibitem{davey:cl98}
M. Davey and D.J.C. MacKay, ``Low Density Parity Check Codes over GF$(q)$,''
{\em IEEE Comm. Lett.}, vol. 2, pp. 165-167, June 1998.

\bibitem{Sridhara2002}
D. Sridhara and T.E. Fuja, ``Low Density parity Check Codes over Groups and Rings,''
{\em in proc. of ITW'02}, Bangladore, India, Oct. 2002.

\bibitem{Ryan}
G. Liva, W.E. Ryan and M. Chiani, ``Design of quasi-cyclic Tanner Codes with Low Error Floors,''
{\em in IEEE Intern. Symp. on Turbo Codes}, Munich, April 2006.

\end{itwreferences}

\end{document}